# ¹⁴N Hyperfine and nuclear interactions of axial and basal NV centers in 4H-SiC: A high frequency (94 GHz) ENDOR study 🔓 FREE

F. F. Murzakhanov; M. A. Sadovnikova; G. V. Mamin; S. S. Nagalyuk; H. J. von Bardeleben 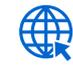 ; W. G. Schmidt 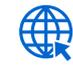 ; T. Biktagirov ✉; U. Gerstmann; V. A. Soltamov ✉

View Online     Export Citation     CrossMark

---

## Articles You May Be Interested In

AIP Publishing

# $^{14}$N Hyperfine and nuclear interactions of axial and basal NV centers in 4H-SiC: A high frequency (94 GHz) ENDOR study 

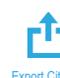

View Online   Export Citation   CrossMark

F. F. Murzakhanov,[1] M. A. Sadovnikova,[1] G. V. Mamin,[1] S. S. Nagalyuk,[2] H. J. von Bardeleben,[3] W. G. Schmidt,[4] T. Biktagirov,[4,a)] U. Gerstmann,[4] and V. A. Soltamov[5,a)]

### AFFILIATIONS

[1]Kazan Federal University, Kazan 420008, Russia
[2]Ioffe Institute, Polytechnicheskaya 26, St. Petersburg 194021, Russia
[3]Sorbonne Universite, Campus Pierre et Marie Curie, Institut des Nanosciences de Paris, 4, place Jussieu, Paris 75005, France
[4]Lehrstuhl für Theoretische Materialphysik, Universität Paderborn, Paderborn 33098, Germany
[5]3rd Institute of Physics, University of Stuttgart, Pfaffenwaldring 57, Stuttgart D-70569, Germany

a)Authors to whom correspondence should be addressed: timur.biktagirov@upb.de and victrosoltamov@gmail.com

## ABSTRACT

The nitrogen-vacancy (NV) centers ($N_C V_{Si}$)$^-$ in $4H$ silicon carbide (SiC) constitute an ensemble of spin $S = 1$ solid state qubits interacting with the surrounding $^{14}$N and $^{29}$Si nuclei. As quantum applications based on a polarization transfer from the electron spin to the nuclei require the knowledge of the electron–nuclear interaction parameters, we have used high-frequency (94 GHz) electron–nuclear double resonance spectroscopy combined with first-principles density functional theory to investigate the hyperfine and nuclear quadrupole interactions of the basal and axial NV centers. We observed that the four inequivalent NV configurations ($hk$, $kh$, $hh$, and $kk$) exhibit different electron–nuclear interaction parameters, suggesting that each NV center may act as a separate optically addressable qubit. Finally, we rationalized the observed differences in terms of distinctions in the local atomic structures of the NV configurations. Thus, our results provide the basic knowledge for an extension of quantum protocols involving the $^{14}$N nuclear spin.

## I. INTRODUCTION

Solid state qubits are now recognized as major elements for numerous applications in quantum technology. In particular, the negatively charged nitrogen-vacancy (NV) defect in diamond and its variants[1–4] and, more recently, vacancy related centers in SiC gave rise to the development of quantum technologies based on solid-state spin defects.[5,6] Quantum sensing, quantum information processing, and quantum memory nodes are only a few examples.[2,7–11] The coherent coupling of the electron spin to the nuclear spin of neighboring nuclei is an important property for some of these applications and is quantified by the hyperfine interaction (HFI) parameters. It has been shown that spin centers in SiC have major advantages over diamond[5,6] linked to the material properties of SiC: a shift of the spectral region of the zero phonon photoluminescence lines to the near-infrared (NIR) for divacancies (VV) and nitrogen vacancy (NV) centers (telecommunication O-band), a modified point symmetry in 4H-SiC and 6H-SiC allowing simplified sensing protocols with axial NV centers, mature nanostructuring technologies for the formation of optical cavities, nanopillar formation, and waveguide incorporation.[12–19] The latter is especially the case for the 4H-SiC polytype, which is widely used in semiconductor technology. Both silicon vacancy $V_{Si}$[20–25] and silicon vacancy related defects such as VV[26–28] and NV in SiC[12–19] have been investigated in detail, via ensemble and single-defect spectroscopy.[3,13] These centers show unique properties, such as ultra-long spin-relaxation times,[15,16,22,26] high Debye–Waller factors, and narrow optical transitions in the near-infrared range.[14,23,27] They can thus be used as spin-photon interfaces, quantum sensors, coherent microwave (MW) amplifiers, and scalable quantum memory nodes to cite some.[7–11,24,25] In this work, we focus on the negatively charged spin $S = 1$ NV centers in 4H-SiC and their interaction with the surrounding $^{14}$N nuclei.

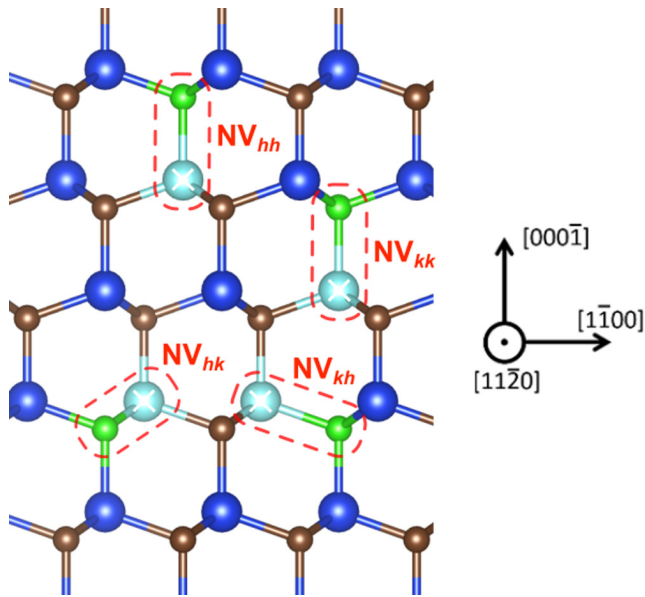

**FIG. 1.** Axial ($NV_{kk}$, $NV_{hh}$) and basal ($NV_{hk}$, $NV_{kh}$) NV defect configurations in 4H-SiC. Silicon vacancies denoted with crosses. Silicon and carbon atoms are shown with blue and brown colors, respectively. Nitrogen atoms are shown in green; crystallographic directions are indicated with arrows.

The crystal structure of 4H-SiC gives rise to four different NV center configurations as shown in Fig. 1. The "family" of NV centers are pairs of a negatively charged silicon vacancy and a nitrogen atom on a nearest neighbor carbon site. The non-equivalent Si lattice sites, quasicubic (k) or hexagonal (h) and the axial or basal N location, give rise to the four different NV defect configurations, denoted as $NV_{kk}$, $NV_{hh}$, $NV_{hk}$, and $NV_{kh}$, with either axial $C_{3v}$ symmetry (kk,hh) or lower $C_{1h}$ symmetry (hk,kh). We have recently determined the $^{14}$N hyperfine interaction (HFI) and nuclear quadrupole interaction (NQI) of one of these centers, $NV_{kk}$, by means of electron–nuclear double resonance (ENDOR) and density functional theory (DFT),[29] In this work, we extend to the other three NV center configurations and analyze the difference in their spin properties from the perspective of qubit application.

We then dispose of the complete set of hyperfine interactions with the central $^{14}$N nuclear spin ($I = 1$) for the four NV centers in 4H-SiC. In the previous studies,[12–14] based on CW ESR spectroscopy, only absolute values of the hyperfine interaction were reported. Furthermore, the weakness of the interaction ($\approx$1.2 MHz) did not allow to reveal a potential anisotropy of this interaction. However, high frequency electron–nuclear double resonance spectroscopy is able to address this issue, since nuclear magnetic resonance (NMR) transitions are measured with kHz-resolution.

## II. EXPERIMENTAL METHOD

The sample used in this study was cut from a commercial n-type, (0001) face 4H-SiC bulk crystal with a nitrogen concentration of $2 \times 10^{17}$ cm$^{-3}$. Sample dimensions were $0.8 \times 0.4 \times 0.2$ mm$^3$. The NV centers were generated in a two step process: first, irradiation with high-energy (12 MeV) protons at a fluence of $1 \times 10^{16}$ cm$^{-2}$, followed by thermal annealing at T = 900 °C. In high temperature annealing, the primary irradiation induced silicon vacancy defects become mobile and a fraction will form complexes with carbon substituted N atoms, which are stable under these conditions. We have previously identified the optimal temperature for NV center creation to be between 700 and 900 K.[12–14]

Electron spin echo-detected (ESE) ESR measurements were performed with a Bruker Elexsys E680 spectrometer, operating at W-band frequency ($\nu = 94$ GHz). A standard Hahn-echo microwave (MW) pulse sequence, $\frac{\pi}{2} - \tau - \pi - ESE$, where $\frac{\pi}{2} = 40$ ns, $\pi = 80$ ns, and $\tau = 240$ ns, was used for the pulsed ENDOR measurements. The NMR transitions between the nuclear spin sublevels of the NV centers were measured by means of a Mims-pulse sequence, $\frac{\pi}{2} - \tau - \frac{\pi}{2} - T(=80 \mu s) - \frac{\pi}{2} - ESE$, with a radio frequency (RF) $\pi$-pulse applied between the second and third MW pulses.[30] The *EasySpin* software package was used for the analysis of the ESR and ENDOR data.[31]

## III. COMPUTATIONAL METHOD (DFT)

The calculation of $^{14}$N electron–nuclear interactions for the NV centers in 4H-SiC was carried out within density functional theory using the Perdew–Burke–Ernzerhof (PBE) exchange-correlation functional.[32] We used the Quantum ESPRESSO software,[33,34] plane-wave basis set with a kinetic energy cutoff of 50 Ry, and scalar-relativistic norm-conserving pseudopotentials. NV centers were modeled in 432-atom supercells of 4H-SiC. All defect structures were allowed to fully relax using equidistant $2 \times 2 \times 2$ k-point samplings until the forces were less than $10^{-4}$ Ry/bohr.

The electron–nuclear coupling parameters were calculated using the GIPAW module of Quantum ESPRESSO. The nuclear quadrupole interaction parameters are related to the electric-field gradient (EFG) tensor. The latter was calculated as the second derivative of the electrostatic potential at the $^{14}$N nucleus, $V_{ij} = \frac{\partial^2 V}{\partial X_i \partial X_j}$. The resulting NQI tensors of the modeled defect structures were computed by considering the spreading of the nuclear quadrupole moment $Q_N$ between 0.0193 and 0.0208 barn reported in the literature.[35]

As will be shown below, the hyperfine interaction at the central $^{14}$N nuclei is of rather indirect nature, behaves sensitively on tiny structural details, and requires great care with respect to fundamental approximations. Relativistic effects are, thus, taken into account in scalar-relativistic approximation,[36,37] so that the Fermi-contact term (the isotropic HF splitting) is determined in a very small, but finite region around the nucleus, represented by $\delta_T(r)$, a smeared-out $\delta$-function with halfwidth $r_T = Ze^2/mc^2$, about 10 times the nuclear radius. By this, the $^{14}$N HFI tensors are obtained from the calculated electron spin density distribution $n_s(\mathbf{r})$ defined with respect to the position of the nucleus, $A_{ij} = \frac{\gamma_I \gamma_e \hbar^2}{2S} \int d^3r \, n_s(\mathbf{r}) \left[ \left( \frac{8\pi}{3} \delta_T(r) \right) + \left( \frac{3x_i x_j}{r^5} - \frac{\delta_{ij}}{r^3} \right) \right]$, where the first parenthesis describes the scalar-relativistic form of the isotropic Fermi-contact contribution and the second parenthesis corresponds to the anisotropic dipole–dipole term.

## IV. RESULTS AND DISCUSSION

### A. Part I. Axial NV_hh center

In Fig. 2, we show the ESR spectrum observed for the magnetic field applied parallel to the crystal $c$ axis, i.e., the defect symmetry axis (B//c), at T = 150 K under optical excitation with a wavelength of $\lambda = 532$ nm. The spectrum is composed of several silicon vacancy related defects. These defects were previously identified: divacancies (VV$_{hh}$ and VV$_{kk}$)[37] and NV centers (NV$_{hh}$ and NV$_{kk}$).[12–14] Optically induced ground state spin polarization is obtained with this nonresonant excitation giving rise to the preferential population of the $m_S = 0$ sublevel. The spin polarization is directly seen through the phase reversal character of the corresponding ESR transitions as schematically shown in the inset in Fig. 2. The spin-Hamiltonian of these $S = 1$ centers is

$$H = g\mu_B \mathbf{B} \cdot \mathbf{S} + \mathbf{SDS} - g_N\mu_N\mathbf{B} \cdot \mathbf{I} + \mathbf{SAI} + \mathbf{IPI}, \quad (1)$$

where the first and the third terms describe the electron ($Z_e$)–Zeeman and nuclear–Zeeman ($Z_n$) interactions, respectively, the second term represents the zero-field splitting (ZFS) with $D = 3/2D_{zz}$, and the fourth and the fifth terms represent the HFI and NQI, respectively. $\mathbf{B}$ is the external static magnetic field, $\mathbf{S}$ is the electron spin operator with $S = 1$, and $\mathbf{I}$ is the nuclear spin operator with $I = 1$ in the case of interaction with $^{14}$N and $I = 1/2$ for $^{29}$Si and $^{13}$C. The g-tensor reflects the axial symmetry relative to the defect symmetry axis and can be expressed by two principal values, $g_{\parallel}$ and $g_{\perp}$. The HFI and NQI tensors $\mathbf{A}$ and $\mathbf{P}$ are given by $A = a + b(3 \cos^2\theta - 1)$ and quadrupole interaction QI $= P(3 \cos^2\theta - 1)$, respectively. Here, $a$ and $b$ are the isotropic and anisotropic parts of the HFI, respectively. $P = \frac{3eQ_N V_{zz}}{4I(2I-1)} = \frac{3}{4I(2I-1)}C_q$ is related to the electric field gradient (EFG) $V_{zz}$ in the direction of the defect symmetry axis and the nuclear electric quadrupole moment $eQ_N$, and $\theta$ is the angle between $\mathbf{B}$ and the symmetry axis of the HFI and NQI tensors.

Since the ZFS parameters and g factors of the divacancies and the NV defects are close, the ESR spectra of the hexagonal VV and NV centers partially overlap (Fig. 2).

Having established the resonance magnetic fields of the fine structure ESR transitions of the NV$_{hh}$ center, we performed Mims-ENDOR experiments. Here, by monitoring the stimulated spin echo intensity (SSE) at the fixed resonant magnetic field, we measured the NMR frequencies corresponding to the nuclear spin flips (selection rules $\Delta m_S = 0$, $\Delta m_I = 1$) driven by a radio frequency (RF) $\pi$-pulse.

First, we have considered the NMR frequencies expected for the NV$_{hh}$ defect fine-structure component corresponding to the $m_S = 0 \leftrightarrow m_S = +1$ transition measured in the resonant magnetic field $B_l$ (as indicated in Fig. 2). The corresponding nuclear Zeeman, quadrupole, and hyperfine interactions lead to the splitting of the pure triplet states $m_S = 0$ and $m_S = +1$ into six sublevels, as schematically shown in the inset in Fig. 3. A sweep of the RF $\pi$-pulse induces nuclear spin flips as depicted by red bars in the inset, thus giving rise to the four possible NMR transitions. These NMR frequencies are determined by the following equation:

$$\nu = |-g_N\mu_N B + m_s(a + b(3\cos^2\theta - 1)) + m_q P(3\cos^2\theta - 1)|/h, \quad (2)$$

where $m_q = 1/2(m_I + m_I')$ is the average value of the two nuclear spin manifolds between which NMR transitions occur, $\nu_L = g_N\mu_N B/h$ is the nuclear Larmor frequency, and $a$ and $b$ are the isotropic (Fermi contact) and anisotropic parts of the HFI, respectively, whereas $P$ is the NQI parameter.[38] At the given orientation, B//c, the four NMR transition frequencies are, thus, the following: $\nu_{1,3} = \nu_L \pm P$ and $\nu_{2,4} = \nu_L - A_{\parallel} \pm P$, where $A_{\parallel} = a + 2b$ is the hyperfine interaction constant and $P$ is the nuclear quadrupole splitting. These frequencies $\nu_{1-4}$ are shown in the top ENDOR spectrum of Fig. 3. From this result, the absolute value of the quadrupole splitting is derived as half of a difference between the resonance frequencies $\nu_1$ and $\nu_3$. Thus, for the NV$_{hh}$ center, the quadrupole splitting is $P = 1.89$ (5) MHz. The absolute value of the hyperfine interaction constant is derived from the measured ENDOR frequencies as $A_{\parallel} = 1.165$ MHz.

Both NQI and HFI parameters for the NV$_{hh}$ defects are slightly larger than those previously reported for the NV$_{kk}$ centers, that is, $P = 1.81$ MHz and $A_{\parallel} = 1.14$ MHz.[29] The NV$_{kk}$ ENDOR spectrum measured under the same experimental conditions (Fig. 3, bottom) demonstrates this small but measurable difference between $NV_{hh}$ and $NV_{kk}$. Indeed, all $NV_{kk}$ ENDOR frequencies are shifted toward the $\nu_L(^{14}N)$ frequency, indicating that for this defect configuration, hyperfine and quadrupole interactions are characterized by smaller values.

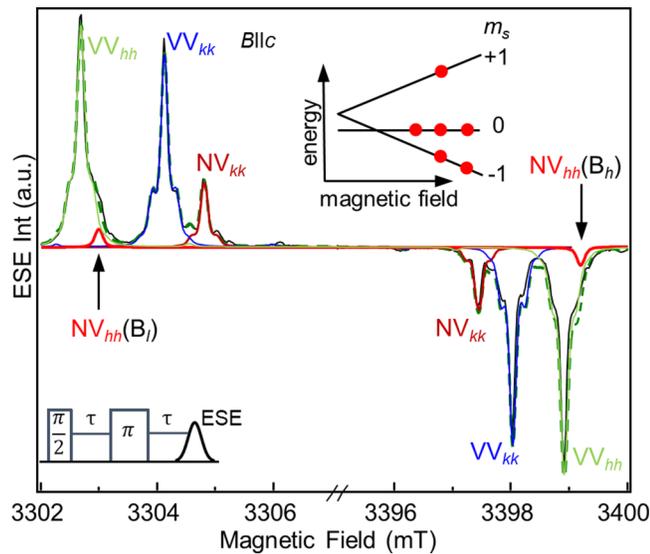

**FIG. 2.** W-band ESR spectrum at a temperature $T = 150$ K registered under $\lambda = 532$ nm excitation. MW pulse sequence used for the detection is shown in the bottom inset. Static magnetic field is applied parallel to the $c$ axis ($\mathbf{B}//c$). The corresponding simulation using Eq. (1) is shown with dashed lines as a superposition of the divacancies (blue and green lines) and NV defect (red and pink lines) ESR signals. Parameters used in the simulation are the following: $D = 1350$ MHz with $g_{\parallel} = 2.0047$ and $D = 1317.5$ MHz with $g_{\parallel} = 2.0046$ for VV$_{hh}$ and VV$_{kk}$, respectively; $D = 1349.5$ MHz with $g_{\parallel} = 2.0046$ and $D = 1299.5$ MHz with $g_{\parallel} = 2.0045$ for NV$_{hh}$ and NV$_{kk}$, respectively. Hyperfine interaction parameters with $^{29}$Si and $^{13}$C shells are taking into account in accordance with Ref. 12 in order to reproduce the ESR line shapes. Transitions of the NV$_{hh}$ defect in the low ($B_l$) and in the high ($B_h$) magnetic fields are shown with vertical arrows. The top inset demonstrates optically induced predominant population of the $m_S = 0$ in the triplet ground state.

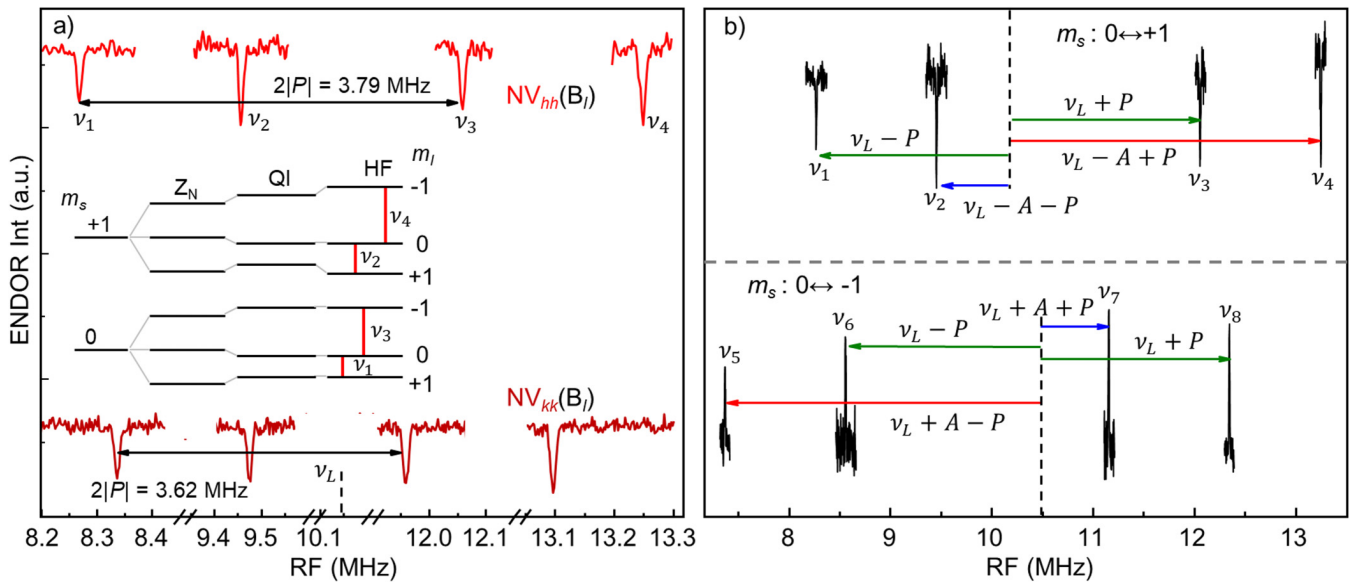

**FIG. 3.** (a) Top spectrum represents the ENDOR data measured in the fixed static magnetic field $B_l$ corresponding to the resonant field of the fine structure transition of the $NV_{hh}$ defects ($m_S = 0 \leftrightarrow m_S = +1$). NMR frequencies $\nu_1 - \nu_4$ are denoted corresponding to the energy level diagram shown in the inset. The bottom spectrum shows ENDOR data collected on the $NV_{kk}$ center's $m_S = 0 \leftrightarrow m_S = +1$ fine structure transition. Horizontal arrows indicate transitions due to the quadrupole splitting in the $m_S = 0$ sublevel. (b) The ENDOR spectra of the $NV_{hh}$ defect depending on the fine structure transitions in the $B_l$ and $B_h$ magnetic fields.

Our previous study of the $NV_{kk}$ defects in $4H$-SiC[29] had established a negative sign of the $^{14}N$ nuclei HFI parameter; this indicates that the spin density distribution is subject of a very specific situation, where the minority spin channel becomes locally dominating. The negative sign of the hyperfine interaction energy with the central nitrogen nucleus is also the case for the NV defect in diamond. The same is found here for the $NV_{hh}$ defect, as directly seen from the ENDOR spectra measured on both the $m_S = 0 \leftrightarrow m_S = +1$ and $m_S = 0 \leftrightarrow m_S = -1$ ESR transitions [see Fig. 3(b)]. The NMR transitions labeled in the ENDOR spectra as $\nu_{2,4}$ and $\nu_{6,7}$ correspond to the nuclear spin flips between hyperfine- and quadrupole-split $m_S = +1$ and $m_S = -1$ Zeeman spin states, respectively. In the $m_S = -1$ state, resonance frequencies are determined by the following equations: $\nu_{6,7} = \nu_L + A \pm P$. Meanwhile in the $m_S = +1$, they are given as $\nu_{2,4} = \nu_L - A \pm P$. That is, if the sign of the hyperfine interaction is negative, one would expect the highest frequency to be detected in the ENDOR spectrum on the $m_S = 0 \leftrightarrow m_S = +1$ fine structure line. As evidenced by the position of the $\nu_4$ resonance in the top spectrum of Fig. 3(b), we conclude that the hyperfine interaction constant for the $NV_{hh}$ defect is negative, with $A_\| = -1.165$ MHz.

The anisotropy of the HF and QI, as measured by the dependence of the ENDOR frequencies related to the $NV_{hh}$ defect on angle $\theta$ between the static external magnetic field $\mathbf{B}$ and the defect symmetry axis (which coincides with the crystal $c$-axis), is presented in Fig. 4. The experimental results are well simulated using Eq. (2), and the parameters deduced are presented in Table I. We have, thus, established the full set of the $NV_{hh}$ spin-Hamiltonian parameters.

We attribute the revealed differences in the electron–nuclear interactions of $NV_{hh}$ and $NV_{kk}$ to differences in the local atomic structures of the defects, as confirmed by our DFT results (see Fig. 5). As shown in Table I, the HFI parameters of both centers are dominated by the isotropic Fermi-contact term $a$. The anisotropic term is found to be small with the $b$ value of the order of 0.01 MHz. The Fermi contact HFI has a negative sign which we explain as follows: the spin density is predominantely localized on the three carbon dangling bonds surrounding the silicon vacancy, which polarizes the core states of the nitrogen. Because of the positive sign of the nuclear magnetic moment of $^{14}N$, the resulting value of $a$ is negative. The small anisotropic contribution is attributed to the dipolar interaction between the electron spin and the nitrogen nucleus at a distance of $\sim 3.4$ Å. This distance is shorter for the $NV_{hh}$ center (cf. Fig. 5), giving rise to the slightly larger HFI values (both $a$ and $b$) observed for this center.

As for the NQI parameters, the observed differences between $NV_{hh}$ and $NV_{kk}$ reflect different bonding characteristics of the $^{14}N$ atoms. We have analyzed these differences with the Townes–Daily model.[39] According to this model, the NQI coupling is related to the occupation $N_z$, $N_x$, and $N_y$ of the nitrogen valence $p_z$, $p_x$, and $p_y$ orbitals. In particular, the $zz$ principal component of the EFG tensor is proportional to the imbalance of $p$-orbital occupations, as given by the relation $V_{zz} \propto (2N_z - N_y - N_x)$.[40,41] By means of a

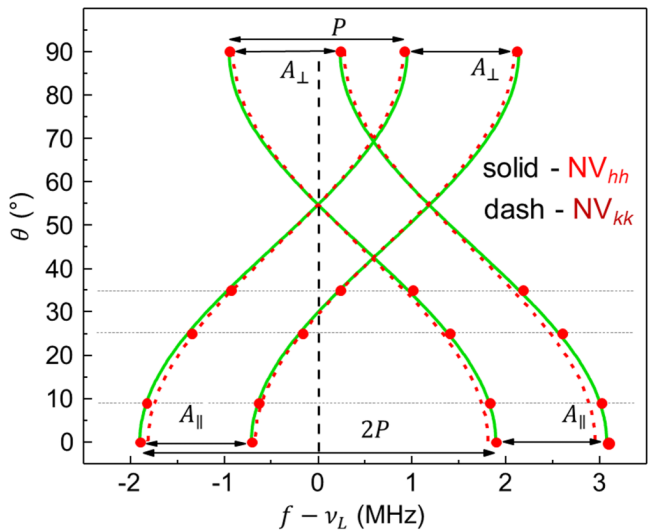

**FIG. 4.** Data points show the angular dependence of ENDOR frequencies for the $NV_{hh}$ defects. Solid curves represent the fit of the data points using Eq. (2). Simulation of the $NV_{kk}$ ENDOR frequencies using parameters $a = -1.170$ MHz, $b = 0.014$ MHz, $P = 1.81$ MHz from Ref. 19 is shown with the dashed curves. Arrows indicate HFI and NQI. The $NV_{hh}$ fitting parameters are $a = -1.185$ MHz, $b = 0.010$ MHz, $P = 1.895$ MHz.

Löwdin analysis,[42] our DFT calculations predict a value of $(2N_z - N_x - N_x) = 0.35e$ for the $NV_{kk}$ center and $0.38e$ for $NV_{hh}$, which agrees with the larger NQI parameter calculated and measured for the latter.

The nitrogen $p_z$ orbital in the NV center contains a lone pair of electrons and is directed toward the vacancy. Accordingly, the three nearest-neighbor carbon atoms of the nitrogen atom are not arranged in a planar structure, but rather in a trigonal pyramidal geometry. Due to electrostatic repulsion between the lone pair and the C–N bonding pair, the relative occupation of the $p_z$ orbital is correlated to the deviation from planarity. From the DFT optimized structures, we find that the C–N–C bond angle in the $NV_{kk}$ center (115.23°) is a little closer to the 120° angle of a trigonal planar geometry compared to $NV_{hh}$ (115.20°). This implies that in the case of $NV_{kk}$, the nitrogen atom is strongly shifted away from the ideal vacancy site along the $c$ axis (cf. Fig. 5). Thus, the results of this simple analysis emphasize the sensitivity of the $^{14}$N NQI parameters to even slight deviations in the atomic structures of the NV centers.

| | | NQI ($^{14}$N) | HFI ($^{14}$N) | |
|---|---|---|---|---|
| | | $eQV_{zz}/\eta$ | $A_{\parallel} = a + 2b$ | $a$ |
| $NV_{kk}$ | Expt.[29] | $|2.41|$ | $-1.142$ | $-1.170$ |
| | DFT | $-2.400 \div -2.226$ | $-1.067$ | $-1.091$ |
| $NV_{hh}$ | Expt. | $|2.53|$ | $-1.165$ | $-1.185$ |
| | DFT | $-2.518 \div -2.336$ | $-1.159$ | $-1.147$ |
| $NV_{kh}$ | Expt. | $|2.43|$ | $-0.970$ | $-1.050$ |
| | DFT | $-2.551 \div -2.367 / 0.027$ | $-0.900$ | $-0.867$ |
| $NV_{hk}$ | Expt. | $|2.31|$ | $-0.650$ | $-0.870$ |
| | DFT | $-2.438 \div -2.264/0.071$ | $-0.617$ | $-0.539$ |

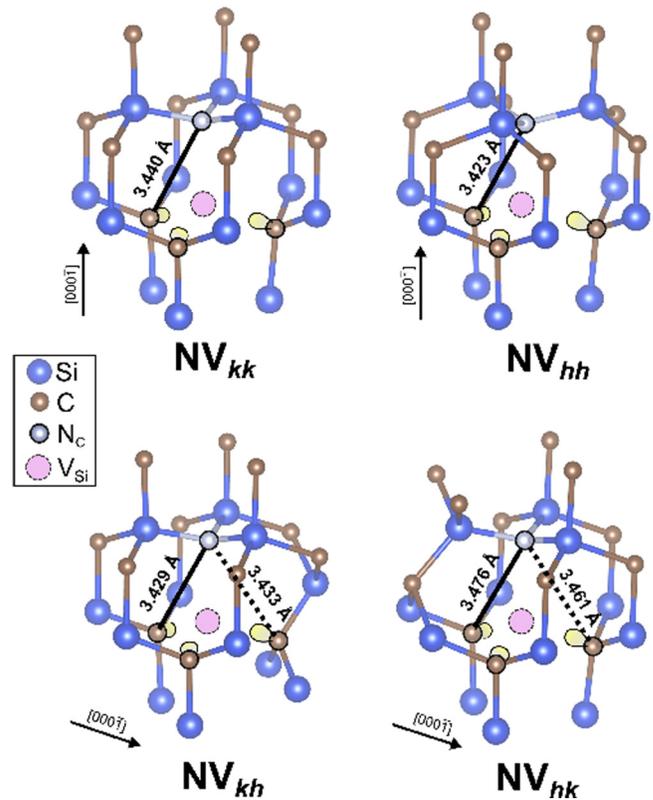

**FIG. 5.** Schematic illustration of the DFT optimized local crystal structures of the $NV_{kk}$, $NV_{hh}$, $NV_{hk}$, and $NV_{kh}$ centers in 4H-SiC (cf. Fig. 1). Yellow shapes at the carbon atoms surrounding the silicon vacancy indicate dangling bonds that constitute the spin density.

### B. Part II. Basal $NV_{kh}$ and $NV_{hk}$ centers

In the second part of this work, we have extended our measurements to the basal $NV_{kh}$ and $NV_{hk}$ centers. The symmetry axes of these basal centers are oriented 70° away from the hexagonal $c$ axis. So, we first rotate the 4H-SiC sample away from the orientation $B \parallel c$ on 70° and measure the optically induced ESR spectrum, as presented in Fig. 6(a). In the following, we will assume that this orientation ($\theta = 70°$) is canonical for basal centers and serves as a parallel direction for $NV_{kh}$ and $NV_{hk}$ centers. In other words, we set this orientation by defining a local angle $\theta' = 0$. The angular dependence of the fine structure components for both basal defects is shown in Fig. 6(b) and is supported by a simulation using spin-Hamiltonian (1), with an additional term $E = 1/2(D_{yy} - D_{xx})$ that reflects the rhombicity of the ZFS. The corresponding parameters are summarized in Table II, and they are close to those previously established for these defects in Refs. 13 and 14 and predicted theoretically in Refs. 13 and 43.

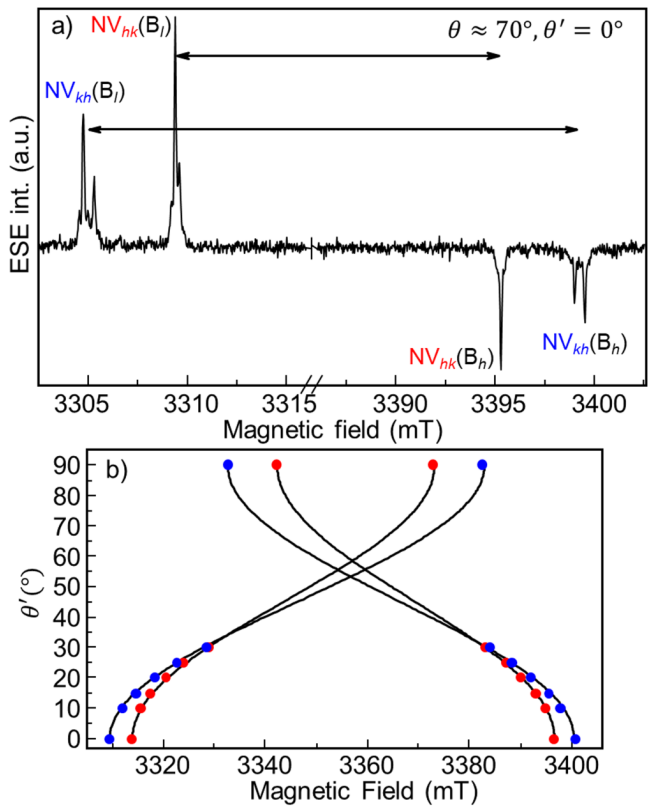

**FIG. 6.** (a) Optically induced ($\lambda = 532$ nm) ESR spectrum of basal NV centers as measured relative to the orientation of the static magnetic field $\boldsymbol{B}$ 70° away from the $c$ axis ($\theta = 70°$, $\theta' = 0$). Fine structure lines ($m_S = 0 \leftrightarrow m_S = +1$) in the low magnetic fields are labeled as $B_l$ and in the high magnetic fields ($m_S = 0 \leftrightarrow m_S = -1$) as $B_h$. (b) Angular variation in the resonance fields of the NV centers for a rotation of the magnetic field in the (11$\overline{2}$0) plane. Here, $\theta'$ is the angle between magnetic field $\boldsymbol{B}$ and the local symmetry axis of the basal NV centers ($\theta'$ is taken to be zero, when $\theta = 70°$).

**TABLE II.** The spin Hamiltonian ESR parameters of basal centers 4H-SiC crystal determined in this work. They are close to previously established for these defects in Ref. 14 and predicted theoretically in Ref. 43.

|  | $g_\perp$ | $g_\parallel$ | $D$ (MHz) | $E$ (MHz) |
|---|---|---|---|---|
| $NV_{kh}$ | 2.0017(1) | 2.0001(1) | 1275(20) | 45(2) |
| $NV_{hk}$ | 2.0015(1) | 2.0000(1) | 1160(20) | 120(5) |

We have analyzed the ENDOR results of the HF and NQ interactions in the same approach as for the axial NV centers in Part I. First, we measured the ENDOR spectra in the canonical orientation for both $NV_{kh}$ and $NV_{hk}$ centers as shown in Fig. 7(a). Significant differences in the HFI and NQI parameters for these defects are seen, reflecting the crystallographic non-equivalence of these defects. The negative sign of the hyperfine interaction is directly deduced for both basal NV defects, since the highest ENDOR frequency observed in the spectra is again $\nu_4$ [see Fig. 7(a)]. We have equally measured the angular dependence of the ENDOR spectra, the results are shown in Fig. 7(b). Fitting the experimental data by Eq. (2) allows to establish the full set of nuclear-interaction

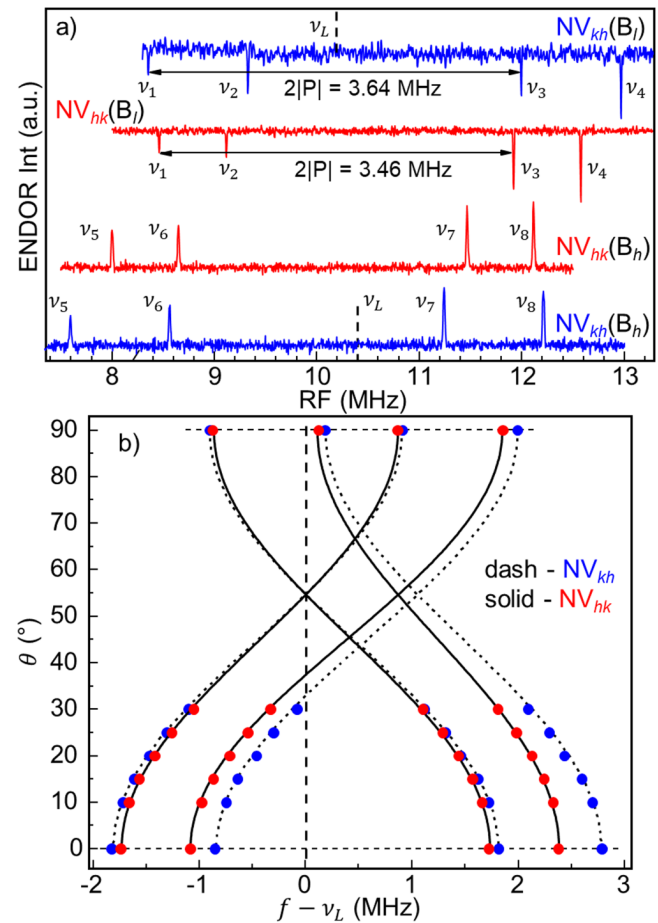

**FIG. 7.** (a) ENDOR spectra of the basal NV centers ($\theta = 70°$ from $c$ axis) measured under $\lambda = 532$ nm excitation. Two top spectra are measured on the $NV_{kh}$ ($B_l$) and $NV_{hk}$ ($B_l$) fine structure lines; bottom spectra are measured on the corresponding high-field ESR lines ($B_h$). The corresponding NMR frequencies are denoted as $\nu_1 - \nu_4$ and $\nu_5 - \nu_8$. (b) Angular dependence of the ENDOR lines for $NV_{kh}$ (blue dots) and $NV_{hk}$ (red dots). Simulations of the dependencies are shown with dashed and solid curves for $NV_{kh}$ and $NV_{hk}$, respectively, using the following parameters: $a = -1.05$ MHz, $b = 0.04$ MHz, $P = 1.82(1)$ for the $NV_{kh}$ defect, and $a = -0.87$ MHz, $b = 0.11$ MHz, $P = 1.73(2)$ MHz for the $NV_{hk}$ defect.

05 October 2023 08:07:58

parameters, the isotropic and anisotropic parts of the HFI, as well as the NQI constants as summarized in the caption of Fig. 7 and in Table I.

While the $^{14}$N HFI and NQI parameters of the basal centers are expected to exhibit deviation from axial symmetry (cf. the ZFS anisotropy parameter $E$ shown in Table II and the DFT optimized defect geometries presented in Fig. 5), we do not resolve this anisotropy in the ENDOR experiment. As for NV$_{kh}$, the DFT results reveal that this center's HFI tensor features almost vanishing in-plane anisotropy, i.e., minor rhombicity (the calculated principal values of the $A$ tensor are $A_{xx} = -0.851$, $A_{yy} = -0.849$, $A_{zz} = -0.900$ MHz). The second basal center, NV$_{kh}$, is characterized by a relatively large rhombic ZFS parameter $E$ (see Table I), indicating that the spin-density distribution at this center deviates significantly from the $C_{3v}$ symmetry. The larger anisotropy of the $^{14}$N HFI agrees with our DFT-calculated principal values of $A_{xx} = -0.523$, $A_{yy} = -0.477$, and $A_{zz} = -0.617$ MHz. The rhombicity of the NQI tensor, expressed by the parameter $\eta$ in Table I, is larger for NV$_{hk}$ compared to NV$_{kh}$. Thus, comparing the different NV configurations, it becomes obvious that small changes in the local defect structure cause discernible changes in the nuclear interaction parameters.

## V. CONCLUSION

With electron–nuclear double resonance spectroscopy, we have determined the complete parameter sets of the ground state spin-Hamiltonians for the four NV centers in 4$H$-SiC. We determined the hyperfine and nuclear quadrupole interaction parameters and rationalized them by DFT calculations: the larger NQI value of the NV$_{hh}$ defect with respect to NV$_{kk}$ is due to a small shift of the $^{14}$N atom away from the vacancy site. The DFT calculations for NV$_{kh}$ and its counterpart demonstrate that even slight differences in the local crystal structure may cause significant changes in the HFI values. Although both NV$_{kh}$ and NV$_{hh}$ feature the substitutional nitrogen at the $h$ site, the HFI of the latter is two times smaller. The present results provide the crucial data for establishing protocols for transferring the electron spin polarization to the surrounding nuclear spins. The latter shows significant promise, particularly due to the recent successful generation of a single NV defect in 4$H$-SiC through ion implantation and subsequent optical readout of the single NV defect spin.[15,19] Moreover, these advancements are coupled with the theoretically predicted exceptionally long coherence times of these defects.[44]

## ACKNOWLEDGMENTS

This work was funded by the subsidy allocated to Kazan Federal University for the state assignment in the sphere of scientific activities (Project No. FZSM-2022-0021). S.S. Nagalyuk acknowledges the support of state assignments of the Ministry of Science and Higher Education of the Russian Federation to Ioffe Institute (No. 0040-2019-0016). The authors acknowledge computing time provided by the Paderborn Center for Parallel Computing (PC2).

## AUTHOR DECLARATIONS

### Conflict of Interest

The authors have no conflicts to disclose.

### Author Contributions

**F. F. Murzakhanov:** Data curation (equal); Formal analysis (equal); Investigation (equal). **M. A. Sadovnikova:** Formal analysis (equal); Investigation (equal); Visualization (equal). **G. V. Mamin:** Data curation (equal); Formal analysis (equal); Investigation (equal); Visualization (equal). **S. S. Nagalyuk:** Formal analysis (equal). **H. J. von Bardeleben:** Conceptualization (equal); Project administration (equal); Writing – review & editing (equal). **W. G. Schmidt:** Conceptualization (equal); Formal analysis (equal); Methodology (equal); Writing – review & editing (equal). **T. Biktagirov:** Formal analysis (equal); Methodology (equal); Software (equal); Visualization (equal); Writing – original draft (equal); Writing – review & editing (equal). **U. Gerstmann:** Formal analysis (equal); Methodology (equal); Software (equal); Visualization (equal); Writing – review & editing (equal). **V. A. Soltamov:** Conceptualization (equal); Formal analysis (equal); Funding acquisition (equal); Supervision (equal); Writing – original draft (equal); Writing – review & editing (equal).

## DATA AVAILABILITY

The data that support the findings of this study are available from the corresponding authors upon reasonable request.